\documentclass[journal]{IEEEtran}
\usepackage{amsthm}
\usepackage{amssymb}
\usepackage{graphicx}
\usepackage{amsmath}
\usepackage{bm}
{}
{}
{}
{}
{}

\usepackage{algorithm}
\usepackage{algpseudocode}
\usepackage{graphics}
\usepackage{epsfig}
\usepackage{multirow}
\usepackage{array}
\usepackage{cite}
\usepackage{stfloats}
\usepackage{subfigure}
\usepackage{booktabs}
\usepackage{color}
\usepackage{tabu} 
\usepackage{stfloats}
\usepackage{mathtools}
\usepackage{stmaryrd} %CP Decomposition, \llbracket \rrbracket 
\begin{document}

\title{Terahertz Integrated Sensing and Communication-Empowered UAVs in 6G: A Transceiver Design Perspective}

\author{\IEEEauthorblockN{
		Ruoyu Zhang,
		Wen Wu,
		Xiaoming Chen, 
		Zhen Gao,
		Yueming Cai}
\thanks{
	This work was supported in part by the National Natural Science Foundation of China under Grants 62201266, 62471036, 62171464, 
	in part by the Natural Science Foundation of Jiangsu Province under Grant BK20210335, 
	in part by the Beijing Natural Science Foundation under Grant L242011, in part by the Shandong Province Natural Science Foundation under Grant ZR2022YQ62, and in part by the Beijing Nova Program. 
    \emph{(Corresponding author: Wen Wu).} }
	\thanks{
		R. Zhang, W. Wu, and Y. Cai are with the 
		Key Laboratory of Near-Range RF Sensing ICs \& Microsystems (NJUST), Ministry of Education, School of Electronic and Optical Engineering, 
		Nanjing University of Science and Technology, Nanjing 210094, China 
		(e-mail: ryzhang19@njust.edu.cn; wuwen@njust.edu.cn; cym23@njust.edu.cn). 
		X. Chen is with the College of Information Science and Electronics Engineering, Zhejiang University, Hangzhou 310027, China (e-mail: chen\_xiaoming@zju.edu.cn).
		Z. Gao is with the School of Information and Electronics, Beijing Institute of Technology, Zhuhai 519088, China (e-mail: gaozhen16@bit.edu.cn).
		}
}

% use for special paper notices
%\IEEEspecialpapernotice{(Invited Paper)}

% make the title area
\maketitle

% As a general rule, do not put math, special symbols or citations
% in the abstract
%\begin{abstract}
%f
%\end{abstract}

%% no keywords
%\begin{IEEEkeywords}
%Sparse vector coding, 
%\end{IEEEkeywords}

% For peer review papers, you can put extra information on the cover
% page as needed:
% \ifCLASSOPTIONpeerreview
% \begin{center} \bfseries EDICS Category: 3-BBND \end{center}
% \fi
%
% For peerreview papers, this IEEEtran command inserts a page break and
% creates the second title. It will be ignored for other modes.
\IEEEpeerreviewmaketitle

\begin{abstract}
Due to their high maneuverability, flexible deployment, and low cost, unmanned aerial vehicles (UAVs) are expected to play a pivotal role in not only communication, but also sensing. 
Especially by exploiting the ultra-wide bandwidth of terahertz (THz) bands, integrated sensing and communication (ISAC)-empowered UAV has been a promising technology of 6G space-air-ground integrated networks.
In this article, we systematically investigate the key techniques and essential obstacles for THz-ISAC-empowered UAV from a transceiver design perspective, with the highlight of its major challenges and key technologies. 
Specifically, we discuss the THz-ISAC-UAV wireless propagation environment, based on which several channel characteristics for communication and sensing are revealed. 
We point out the transceiver payload design peculiarities for THz-ISAC-UAV from the perspective of antenna design, radio frequency front-end, and baseband signal processing.
To deal with the specificities faced by the payload, we shed light on three key technologies, i.e., hybrid beamforming for ultra-massive MIMO-ISAC, power-efficient THz-ISAC waveform design, as well as communication and sensing channel state information acquisition, and extensively elaborate their concepts and key issues. 
More importantly, future research directions and associated open problems are presented, which may unleash the full potential of THz-ISAC-UAV for 6G wireless networks.
\end{abstract}


%Index Terms—Integrated sensing and communication, UAV, ISAC-enabled UAV, UAV-assisted ISAC

%has attracted great attention i
% to pursue direct tradeoffs between them as well as mutual performance gains,
\section{Introduction}
% no \IEEEPARstart

%Background
%Applications

Unmanned aerial vehicles (UAVs), driven by their high maneuverability, flexible deployment, and low cost, are expected to be one of the pivotal components in space-air-ground integrated networks for 6G and beyond \cite{Mozaffari2021Toward}. 
As an emerging aerial platform, UAV can be employed as an airborne base station, an access point, or a relay to enhance the wireless connectivity relying on the communication payload. 
The sensing payload equipped on UAV can enable the ubiquitous localization, detection, and imaging, facilitating the instantaneous perception of the environment and all-weather radar sensing capability.  
The indispensable communication and sensing functionalities support a wide variety of emerging application scenarios including a hotspot area, disaster relief, aerial surveillance, and cargo delivery, among many others \cite{Fei2023AirGround}.

\begin{figure*}[!tp]
	\centering
	\includegraphics[width=6.0in]
	{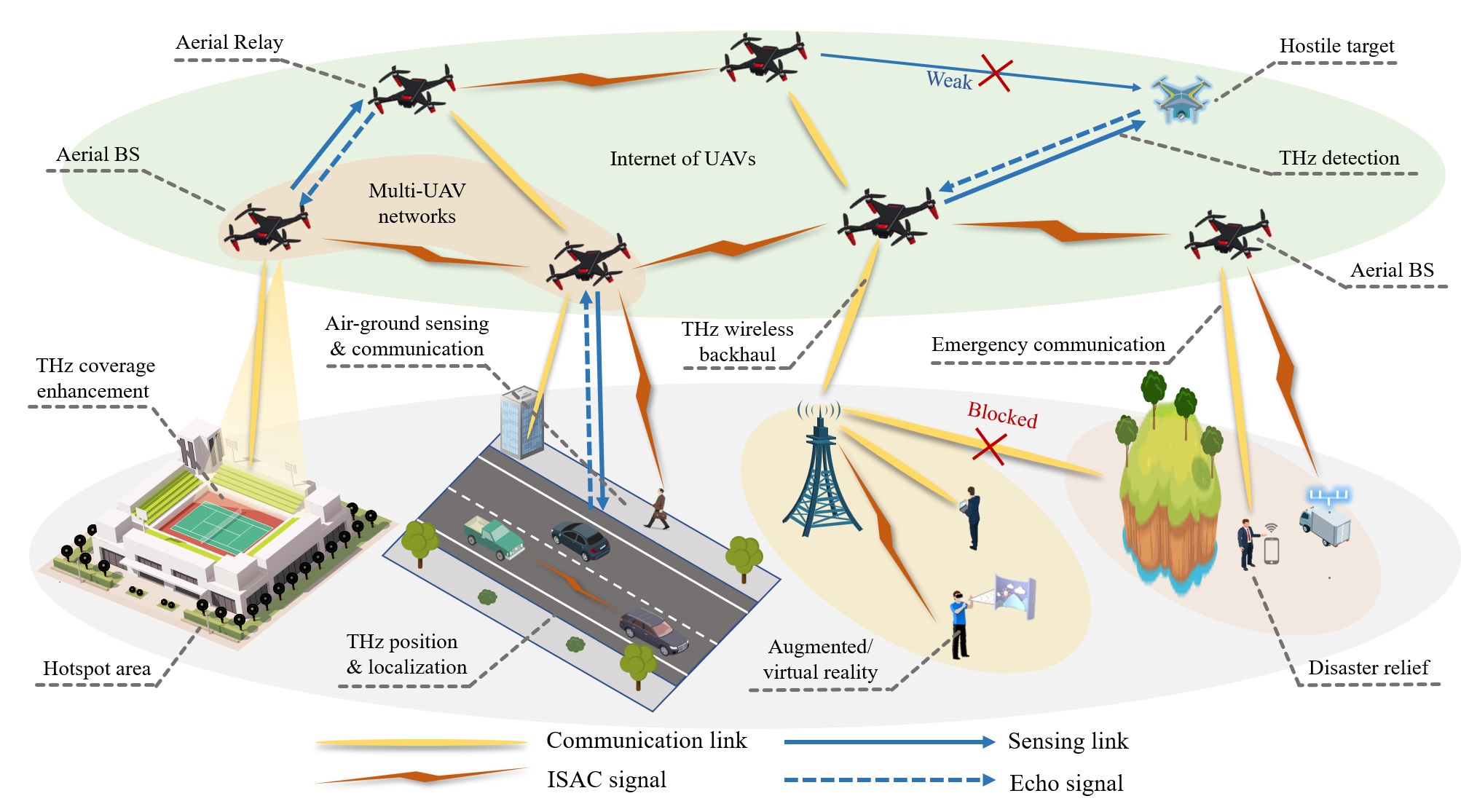}
	\caption{An overview of applications of THz-ISAC-empowered UAV in 6G.}
	\label{Fig_SystemModel}
\end{figure*}

Terahertz integrated sensing and communication (THz-ISAC) has been envisioned as a promising technology to unify the dual functionalities into a single system, opening up new opportunities for enhancing UAV operations \cite{chen2021terahertz,Mu2023UAVISAC}. 
On one hand, benefitting from shared hardware modules, 
both communication and sensing payloads can be replaced by the ISAC payload, thereby reducing the device size, weight, power consumption, and computational complexity. 
On the other hand, due to the extremely small wavelength in THz bands, the size of transceiver tends to be much smaller, which enables a compact, miniaturized, and integrated design of THz-ISAC payloads \cite{Azari2022THzUAVs}.
More importantly, recent progresses in semiconductor industries may facilitate the design of THz transceiver devices, thereby addressing a long-standing barrier ahead the widespread deployment of THz technology \cite{wang2021key}.
In conjunction with the extensive spectrum resources of THz bands, as depicted in Fig. \ref{Fig_SystemModel}, THz-ISAC-empowered UAV (THz-ISAC-UAV) is anticipated to provide novel opportunities for ultra-high communication and sensing applications in 6G.

Despite the attractive merits and thriving prospects, there are still a plethora of new technical challenges and practical transceiver implementation issues remain unsolved for THz-ISAC-UAV.
Firstly, the wireless propagation environment for communication and sensing links can be highly complex at THz bands, which becomes further complicated by comprising the unpredictable movement and rotation of UAV platforms.
Secondly, albeit of the recent advances of the THz circuits and devices, the high power demand of THz-ISAC transceivers causes serious expenditure issues for confined UAV platforms, not to mention the severe round-trip path loss during the target sensing process.
Thirdly, despite the promising prospects of ISAC, the dissonance between two functionalities requires judicious consideration and designs, especially for resource-limited UAV payloads and the power-limited THz devices. 
As summarized in Table \ref{Table1}, existing researches have been conducted on the related technologies of THz, ISAC, and UAV. 
However, the stringent power, weight, and size constraints imposed by UAVs, especially for the widespread commercial miniaturized drones (e.g., DJI, Parrot), make it challenging to accommodate both communication and sensing payloads. 
In order to fundamentally cope with the complex channel environment, multifunctional task requirements, and stringent payload constraints faced by THz-ISAC-UAV, there is an urgent need to emphasize systematic research efforts from the transceiver design perspective.

\begin{table*}[!tp]
	\centering
	\caption{Summary and comparison of existing magazine papers for THz, ISAC, and UAV.}
	\includegraphics[width=7.0in]
	{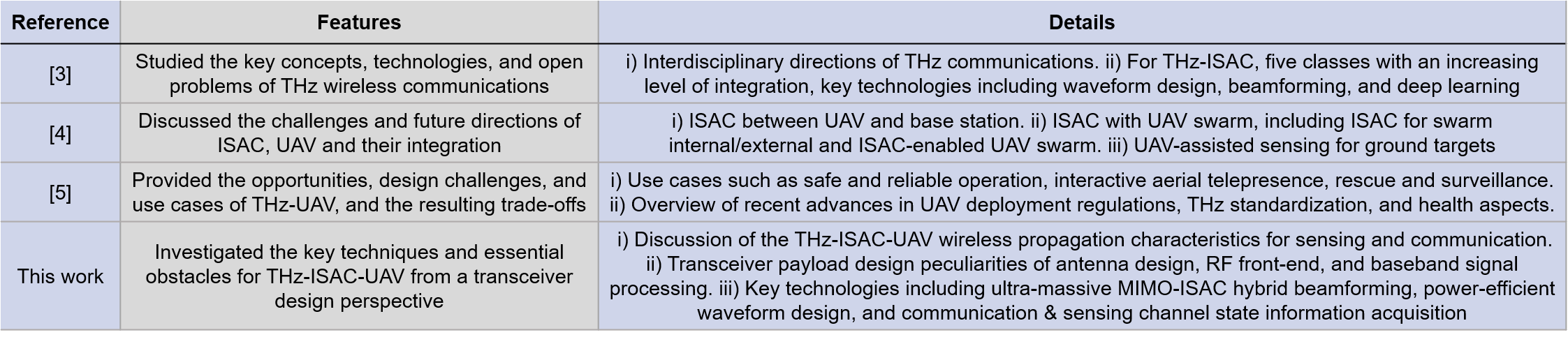}
	\label{Table1}
\end{table*}

In this article, we investigate the key techniques and essential obstacles of transceiver designs for THz-ISAC-UAV, with the emphasis on its challenges and key technologies. 
We first discuss the propagation environment of THz-ISAC-UAV, based on which several channel characteristics for communication and sensing are revealed. 
Then, we specify the transceiver payload design peculiarity for THz-ISAC-UAV from the perspective of antenna design, radio frequency (RF) front-end, and baseband signal processing, and discuss its key challenges.
To deal with the payload design peculiarities caused by severe propagation loss, power limitation, and high-mobility movements, we particularly shed light on three key technologies for THz-ISAC-UAV, i.e., hybrid beamforming for ultra-massive MIMO-ISAC, power-efficient THz-ISAC waveform design, and communication and sensing (C\&S) channel state information acquisition, and extensively elaborate their concepts and key issues.
Finally, potential research directions of THz-ISAC-UAV are identified, which may facilitate its applications for future 6G wireless networks.

\section{Propagation Characteristic of THz-ISAC-UAV}
\label{sec:THzChannelCharac}
%Characterization and Modeling

The study of the propagation characteristic is the prerequisite for the design of THz-ISAC-UAV.  
In the following, we will discuss the distinct THz-ISAC-UAV propagation characteristics that distinguish it from the ground-based and lower-frequency channels.

\subsubsection{Beam-Split Effects} 
To compensate the substantial THz signal attenuation, ultra-massive antenna arrays are generally employed for UAV payloads to establish pencil beams with ultra-high array gains \cite{sarieddeen2021overview}. 
Due to the ultra-wide bandwidth of THz-ISAC signals, the generated narrow beams may disperse and split into surrounding directions at different frequencies, referring to the beam-split effects. 
Despite this characteristic may deteriorate the array gain for both communication and sensing, it can be exploited to enhance the identification of diverse target directions. For instance, by taking its advantage of simultaneously different beam directions at different subcarriers, a controllable wider coverage can be achieved for sensing targets located in multiple directions at once.

\subsubsection{Non-Stationary Environment}
Unlike the conventional terrestrial propagation environment where the channel is time stationary, high-mobility UAV platforms with THz carrier frequency will introduce severe Doppler effects than that with millimeter wave (mmWave) carrier frequency, leading to the significantly reduced channel coherence time. 
Besides, the wind effects cannot be overlooked and cause extra random movement/rotation of UAVs, which generates additional Doppler shifts and constrains the coherence time. 
Under such a circumstance, the channel statistical properties may be non-stationary even within a short period of time, which imposes stringent necessities on robust sensing, low-overhead channel estimation, and reliable communication.

\subsubsection{Target-Dependent Uncertainty}
The sensing process is essentially to acquire information about the targets of interest from the propagation environment, the THz signals for sensing targets exhibit the round-trip propagation loss and Doppler shifts, leading to even lower signal-to-noise ratio (SNR). 
Extra uncertainties interrelated to sensing targets, such as the target size, distance away from the transceiver, radar cross-section (RCS) related to the material, etc., will be introduced into channel propagation. In particular, targets in the THz bands are more likely to become extended targets in angle and range domains, thereby introducing multiple propagation paths/signal echoes that disperse multiple different beamwidths and time delays/range cells.

\begin{figure}[!tp]
	\centering
	\includegraphics[width=3.2in]
	{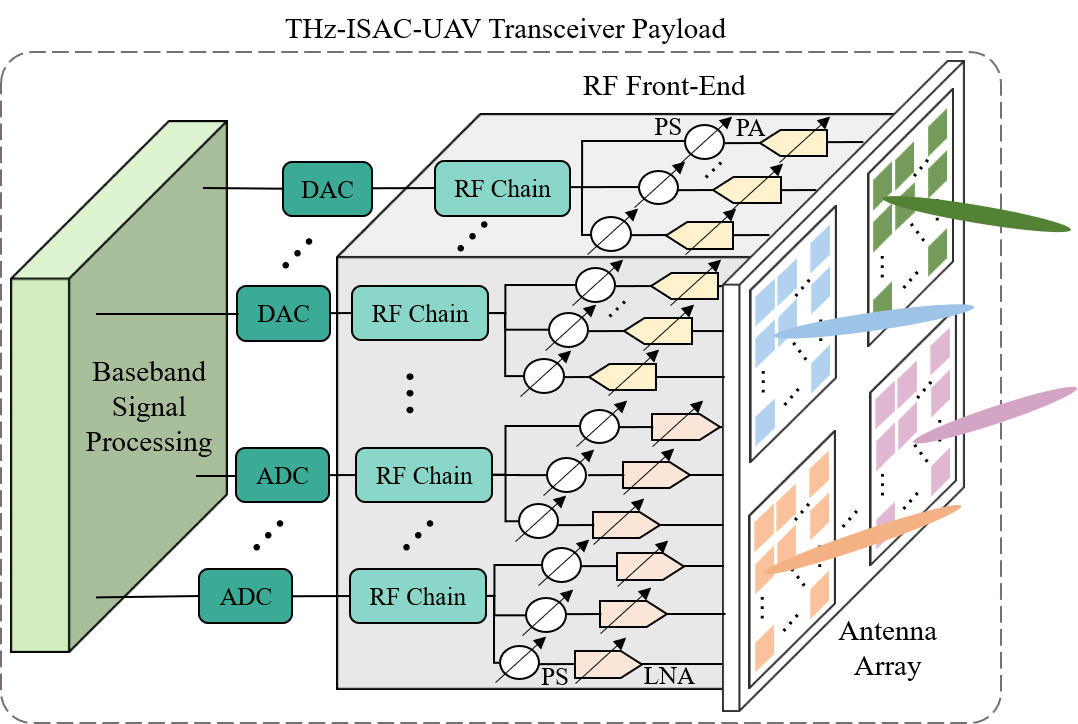}
	\caption{The relationship among the modules of THz-ISAC-UAV transceiver payloads.}
	\label{Fig2}
\end{figure}

\section{Transceiver Payload Design of THz-ISAC-UAV}

Attributed to the uniqueness of the THz-ISAC-UAV propagation environment, as well as the limitations in size, weight, and power consumption, the transceiver payload design has significantly different requirements as compared to the traditional cellular communication and sensing.
In the following, we will discuss the design peculiarity from the perspective of antenna design, RF front-end, and baseband signal processing, respectively. 
The relationship among different modules of THz-ISAC-UAV transceiver payloads is illustrated in Fig. \ref{Fig2}.

\subsection{Antenna Design of THz-ISAC-UAV}
Antenna design is a critical part of THz-ISAC-UAV transceiver payloads, and its electrical characteristics directly affect the performance of information transmission and target sensing. 
To enable the THz wireless links, antenna arrays with sufficiently large gain and high directional capability are required to enhance the equivalent isotropic radiated power. 
The existing works have exploited different operation principles for antenna design, including photoconductive, metallic, and dielectric, etc \cite{wang2021key}. 
Profiting by the recent advancements in microfabrication technologies, researchers are exploring various new materials for antenna and array designs, such as graphene, liquid crystals, and metamaterials, to achieve efficient THz radiation.  
Particularly, by introducing the metasurface with sub-wavelength structures, it becomes possible to integrate an excessive number of low-cost elements, which is beneficial for THz-ISAC-UAV to achieve programmable wireless environments with a hardware-efficient manner.

Another key factor of THz-ISAC-UAV antenna design is to support full-duplex operation within a compact space. Unlike traditional single-function communication systems, ISAC requires receive antennas to capture echo signals for sensing, which makes it crucial to minimize self-interference from transmit antennas. Sufficient isolation between the transmit and receive antennas is necessary to prevent receiver saturation and maintain the dynamic range needed to capture target reflections. Therefore, the antenna design must ensure adequate isolation without compromising the reception of useful target reflections.

\subsection{RF Front-End Design of THz-ISAC-UAV}

As a key component of current THz wireless systems, the RF front-end, along with its architecture and functionality, directly affect the performance of THz-ISAC payloads. 
Power amplifiers (PAs), low noise amplifiers (LNAs), etc., are typical components of RF front-ends, whose performance are closely dependent on breakthroughs of solid-state circuits. 
However, due to the immature III-V compound semiconductor transistor technology in THz bands, the performance of high-efficiency PAs and high-sensitivity LNAs is inherently constrained, which will restrict the communication transmission and target detection distance of THz-ISAC-UAV.

In recent years, as the maximum oscillation frequency of complementary metal oxide semiconductor and SiGe transistors grows, it becomes possible for THz circuits design using the mainstream silicon-based technology. 
This drives the use of integrated circuits (ICs) for THz RF front-end design, which can integrate active circuits and electromagnetic fields from multiple disciplinary perspectives. 
For UAV payloads with the compact space, despite RF front-end modules such as LNA and PA are necessary for both sensing and communication, their performance requirements including signal reception, radiation, or amplification, may be distinct. 
When integrating them into one single system, it is necessary to take their performance differences, reconfigurable functionality switching, etc., into account. 
To this end, ICs-based RF front-end design is an enabling approach, which is promising for fully-integrated THz on-chip microsystems, sharing components, and realizing multi-functional characteristics.

\subsection{Baseband Signal Processing of THz-ISAC-UAV}
%\subsubsection{Challenges}
Baseband signal processing plays an essential role to generate the suitable THz-ISAC waveform, decode the communication information, and acquire the target parameter information \cite{sarieddeen2021overview}. 
Despite the benefits from the advanced antenna design and ICs-based RF front-end design, the dissonance between communication and sensing functionalities must be carefully considered, especially for resource-limited UAV payloads and the power-limited THz RF devices. 
In addition, the movements, irregular trajectory, or unpredictable shaking of UAVs, may deteriorate the THz-ISAC performance, such as more frequent channel estimation, significant delays, and reduced target detection reliability. 

To address the aforementioned issues, advanced signal processing techniques require special investigation, in which the properties of THz ultra-massive MIMO (UM-MIMO) transceiver architectures, the joint design and optimization of THz-ISAC waveforms, as well as the channel and target parameters acquisition in high-mobility UAV environments, can be exploited.
For example, the power-efficient ISAC waveforms that are capable of sharing the same signaling resources can be designed to address the C\&S dissonance, such that the integration gain of ISAC can be achieved. 
By exploiting the ultra-narrow beams and flexible beamforming of UM-MIMO, the information transmission and target sensing at different directions can be simultaneously satisfied under the desired power requirements.
In addition, benefitting from the millimeter-level of THz sensing accuracy and the correlation of C\&S channels, the sensing feedback can be used to help UAV adjust its beam direction, such that the pilot overhead and communication delay can be reduced.

\begin{figure*}[!tp]
	\centering
	\includegraphics[width=6.5in]
	{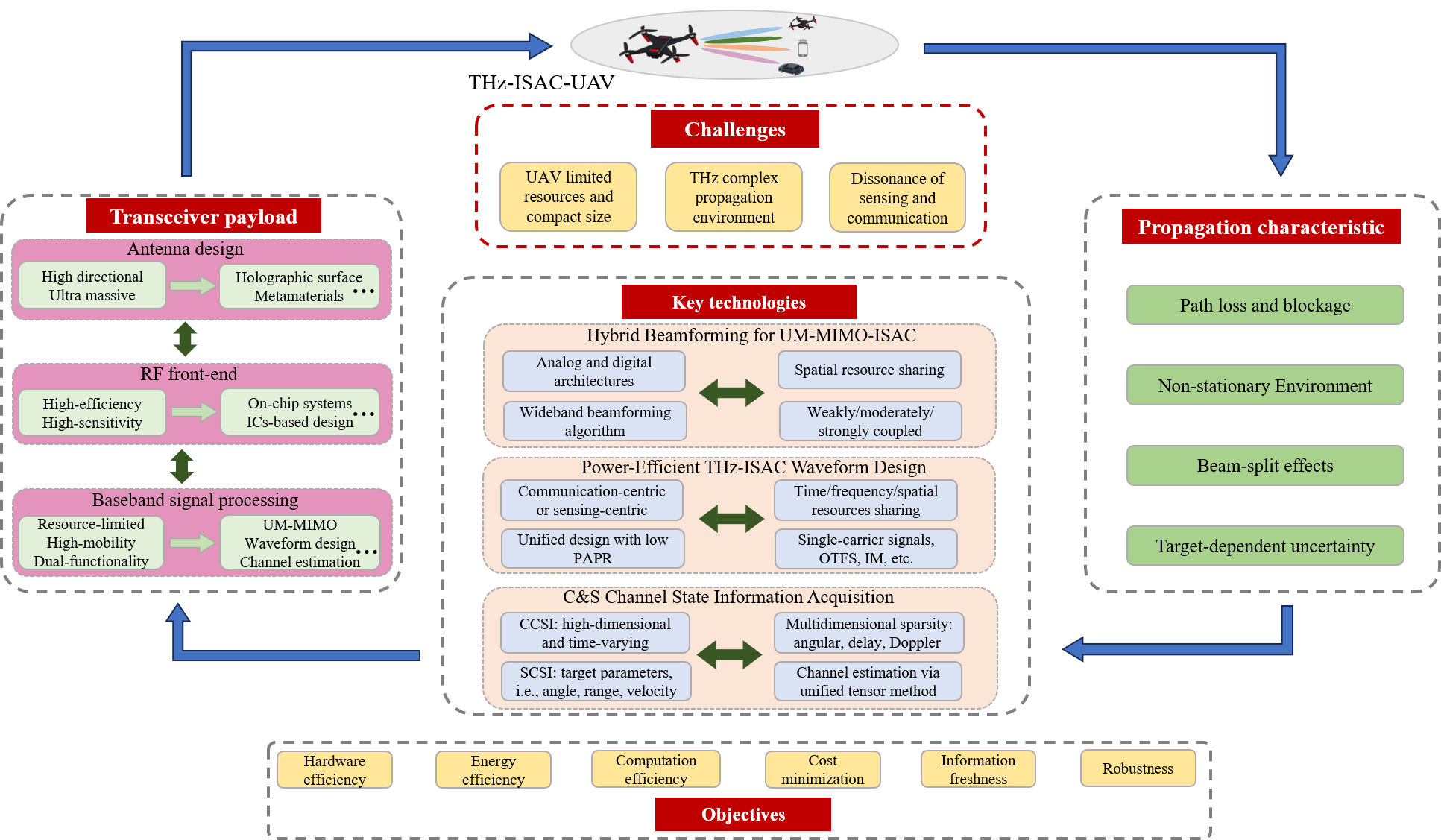}
	\caption{Key techniques of THz-ISAC-UAV for harvesting the peculiarities of propagation characteristics and transceiver payload.}
	\label{Fig5}
\end{figure*}

\section{Key Techniques of THz-ISAC-UAV}

To fully harvest the peculiarities of THz-ISAC-UAV propagation characteristics and transceiver payload, 
we will discuss key techniques for THz-ISAC-UAV in this section, including hybrid beamforming for UM-MIMO-ISAC, power-efficient THz-ISAC waveform design, and C\&S channel state information acquisition.
It is worth mentioning that the aforementioned technologies are fundamentally momentous for enabling THz-ISAC-UAV, as depicted in Fig. \ref{Fig5}.

\subsection{Hybrid Beamforming for UM-MIMO-ISAC}

\subsubsection{Concepts} 
To surpass the distance limitations of THz communication and sensing, it is essential to employ UM-MIMO equipped with thousands of antennas for unleashing the true powers of THz bands. 
Traditionally fully-digital UM-MIMO architectures require a dedicated RF chain for each individual antenna element, which makes the hardware complexity prohibitively unbearable for UAV platforms with limited payload space, volume, and power. 
Hybrid beamforming, which combines analog and digital domain signal processing, is an appealing hardware-efficient technique to substantially reduce the required number of RF chains while achieving comparable performance \cite{tan2022thz}. 
As such, it can be expected to provide an effective compromise of implementation complexity and THz-ISAC-UAV performance.

\subsubsection{Architectures}

As discussed in Section \ref{sec:THzChannelCharac}, conventional UM-MIMO with hybrid beamforming architectures suffer from beam-split effects and cause additional array gain loss.
The essential reason is due to the fact that phase shifters (PSs) is only capable of generating frequency-independent weights and phases. 
The use of true-time-delay (TTD) can help to deal with this limitation by generating the frequency-proportional beamforming weights.
For instance, the authors in \cite{tan2022thz} propose a delay-phase-based hybrid beamforming architecture, in which an additional TTD layer is inserted between the RF chains and the PSs network to fulfil the frequency-dependent hybrid beamforming.
This architecture also provides an energy-efficient hardware solution with multi-beam and wideband sensing capability, facilitating the implementation of UM-MIMO for sensing and communication.

\begin{figure*}[!tp]
	\centering
	\includegraphics[width=6.3in]
	{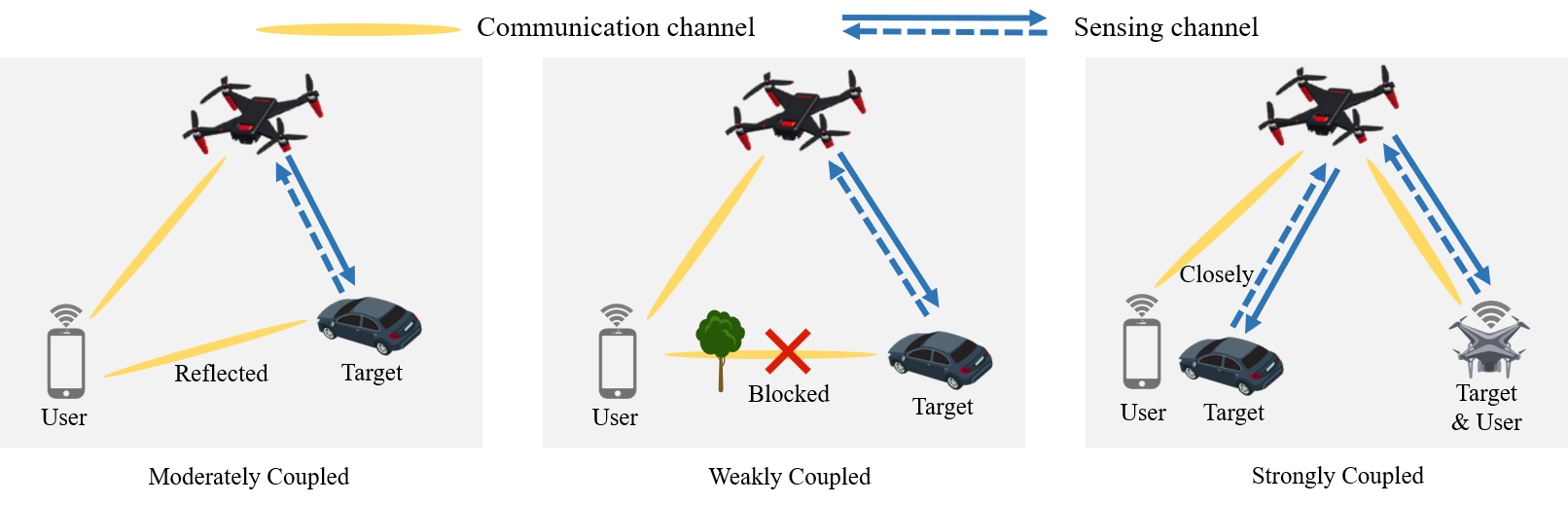}
	\caption{An illustrative example of three coupling cases of communication channels and sensing channels.}
	\label{Fig4}
\end{figure*}

\subsubsection{Algorithms}

Efficient hybrid beamforming algorithms are required to coordinate the objective conflicts of communication and sensing for THz-ISAC-UAV transceiver payloads.
In a typical THz-ISAC-UAV scenario, the sensing functionality tends to exploit the time-varying scan beams or robust beampattern to probe the potential targets, while communication transmission requires a stably directional beam towards the user equipment.  
The performance of beamforming algorithms in THz-ISAC-UAV is also affected by the intricate interrelationship between communication and sensing channels, which can be categorized into three coupling types according to the geometric relationships among users and targets \cite{Chepuri2023Integrated}. 
As shown in Fig. \ref{Fig4}, in the weakly coupled case, the communication user and the sensing target are spatially isolated, where the subspaces of the C\&S channels are nearly orthogonal. 
The moderately coupled case means there exists a partial correlation between C\&S channels, where the beamforming gain for communication and sensing can be partially reused. 
As for the strongly coupled case, the two subspaces of C\&S channels are almost aligned to each other. Accordingly, the communication and sensing functionalities can share the full degrees of freedom of the UM-MIMO entire array, thereby potentially achieving the maximum ISAC performance.

\subsection{Power-Efficient THz-ISAC Waveform Design}
%Beamforming Design
%Waveform Design
%OTFS
\subsubsection{Concepts}
In light of the fact that the communication and sensing performance is usually assessed by different metrics, designing THz-ISAC waveforms that satisfy the requirements of both functionalities is critical for UAV payloads. 
Straightforwardly, we can design the THz-ISAC waveforms such that C\&S signals are orthogonal in temporal, spectral, or spatial domains and do not interfere with each other. 
To further improve the resource efficiency, one can design the fully unified THz-ISAC waveforms and leverage one shared signal to achieve the dual functionalities. 
Based on the underlying signal characteristics and task priorities, we will discuss the communication-centric and sensing-centric waveform designs for the resource-constraint UAV platforms as follows.

\subsubsection{Communication-Centric Waveform Design}
The goal of this design is to leverage the existing communication waveforms to extract target information and enable the sensing functionality. 
Typical communication-based ISAC waveforms can be designed by leveraging the single-carrier and orthogonal frequency-division multiplexing (OFDM) signals. 
Due to the compatibility with current cellular wireless networks, as well as high-efficiency and robust communication transmission, 
OFDM has also been shown flexible sensing capability to enable the perceptive mobile networks \cite{Zhang2021Perceptive}.
Nevertheless, the use of such waveforms inevitably introduces the peak-to-average power ratio (PAPR) issue, which may cause additional power back-off of THz RF front-ends and limit the communication and sensing distances of UAV platforms.
The single-carrier signals, such as the discrete Fourier transform spread OFDM, generally enjoy lower PAPR, but the main disadvantage is the limited data transmission rate. 
Recently, the emerging orthogonal time frequency space (OTFS) waveforms, which employ delay-Doppler domain modulation to encode information, can better resist the impact of high Doppler frequency shift and possess the potential to achieve full diversity gain, thereby achieving reliable UAV communication in high-mobility scenarios \cite{Wei2021Orthogonal}.

\subsubsection{Sensing-Centric Waveform Design}
In this design, the communication information bits can be embedded into existing known sensing waveforms, while maintaining the near-optimal sensing performance.
Since the existing sensing signals generally exhibit relatively constant envelope, this type of design can benefit from inheriting high power efficiency, which is particularly suitable for THz bands. 
For example, the chirp signal, which has been adopted in numerous radar applications, can be applied as carriers and straightforwardly incorporate the communication functionality in combination with modulation, e.g., phase-shift keying.
To promote the communication data transmission rate while guaranteeing the sensing performance, the index modulation (IM) has been recently applied to design sensing-centric ISAC waveforms, where the additional information bits can be conveyed via the indices of resource blocks, such as different antennas, frequencies, and time slots. In this way, the IM-based ISAC waveforms inherit the low PAPR and superior performance of original radar sensing waveform, while improving the spectral efficiency.

\subsection{C\&S Channel State Information Acquisition}

\subsubsection{Concepts} 
Compared to traditional pure communication systems, accurate channel state information is particularly important for both communication and sensing functionalities in THz-ISAC-UAV. 
On one hand, communication channel state information (CCSI) is a prerequisite for beamforming design, data decoding, etc. 
On the other hand, sensing channel state information (SCSI) involves the desired targets parameter information that needs to be extracted from the echo signals. 
However, in the THz frequency band, the high-dimensional C\&S channels resulting from large bandwidth and number of antennas impose significant challenges for THz-ISAC-UAV. 
The extremely narrow beam generated by UM-MIMO also leads to a sharp increase of the required number of beams, incurring the considerable overhead cost for target search.

\begin{figure*}[htbp]
	%\centering	%\label{Fig_AF}
	\subfigure[]{
		\begin{minipage}[t]{0.33\linewidth}
			\centering
			\includegraphics[width=2.4in]{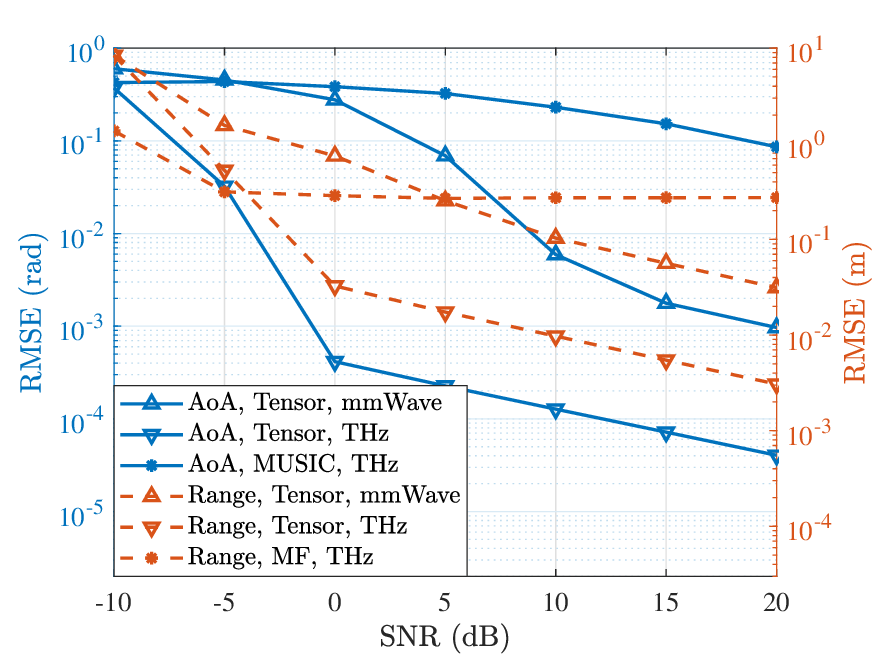}
			%\caption{fig1}
		\end{minipage}%
	}%
	\subfigure[]{
		\begin{minipage}[t]{0.33\linewidth}
			\centering
			\includegraphics[width=2.4in]{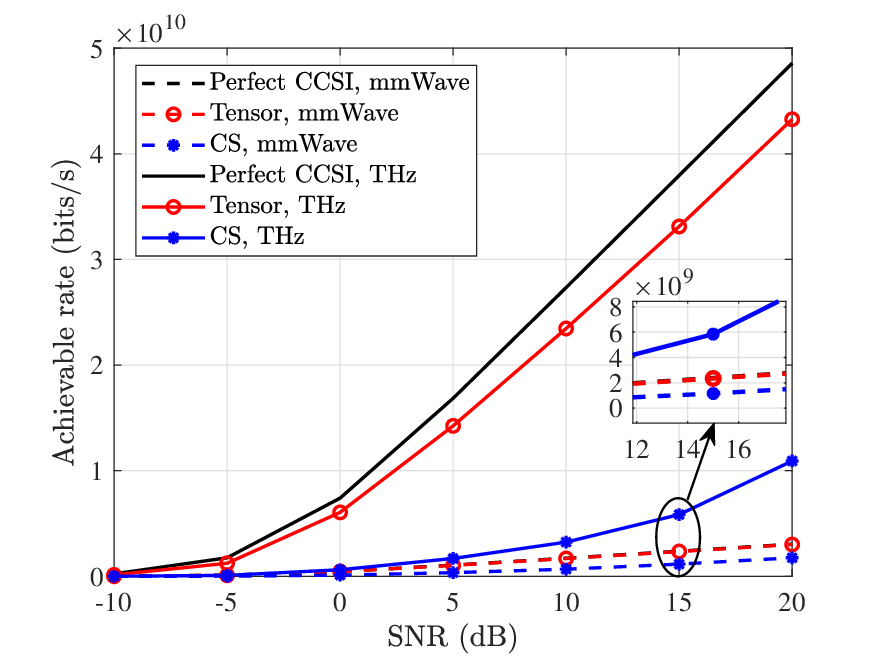}
			%\caption{fig2}
		\end{minipage}%
	}%
	\subfigure[]{
		\begin{minipage}[t]{0.33\linewidth}
			\centering
			\includegraphics[width=2.4in]{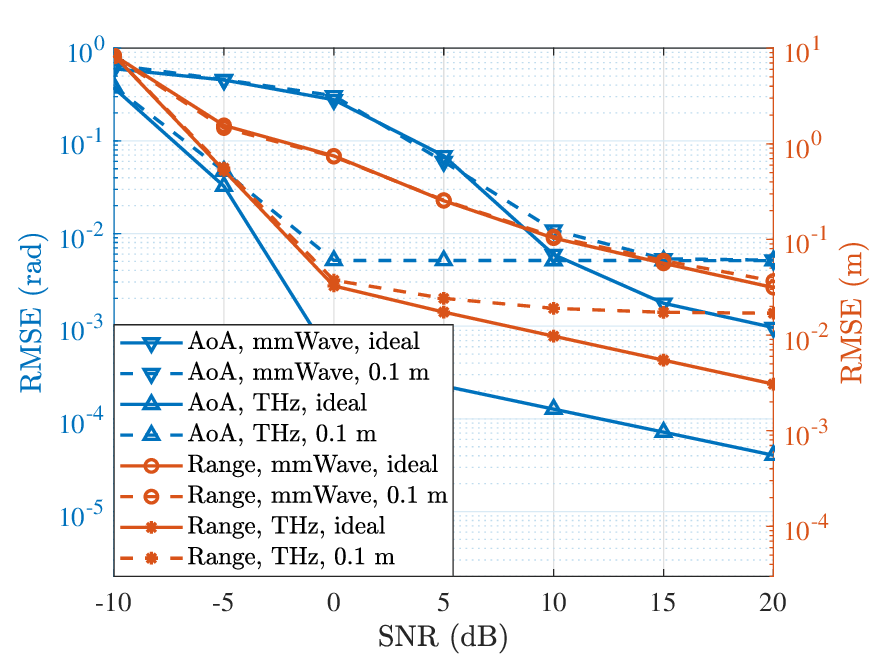}
			%\caption{fig1}
		\end{minipage}%
	}%
	\centering
	\caption{Sensing and communication performance of the THz-ISAC-UAV, where conventional parameter and chanenl estimation methods, and the ISAC performance at mmWave bands, are compared. a) RMSE performance of target sensing; b) communication achievable rate performance; c) RMSE performance of target sensing in the presence of UAV location uncertainty.}
\end{figure*}

\subsubsection{Efficient C\&S Channel Estimation}
In spite of the aforementioned major challenges, the inherent connection between CCSI and SCSI may facilitate the effective channel estimation in THz-ISAC-UAV. 
From communication perspective, the acquisition of CCSI exploits the known pilot signals at both the transmitter and receiver to estimate the propagation environment. 
Similarly, the acquisition of SCSI relies on the active radiation of the known probing signals to detect potential targets in the surrounding environment, where the target parameter information, including distance, velocity, azimuth, RCS, etc., can be obtained.
From this point of view, the process of SCSI acquisition is more similar to that of communication channel estimation, instead of the data transmission itself. 
Within the time duration of coherent processing interval, due to the significantly limited THz scattering environment and ultra-massive antenna array, 
the communication channel will display the three-dimensional sparsity in the delay-Doppler-angular domain when wideband ISAC signals are transmitted.
The high dimensional CCSI can be parameterized by a small number of physical parameters, e.g., angles, path gains, which can establish an intrinsic link to the azimuth angles, RCS, of the target parameter information in SCSI. 
For example, the authors in \cite{Zhang2024TWC} proposed an efficient  method for CCSI \& SCSI acquisition problem in a unified tensor framework, which shows superiority in terms of estimation accuracy, sensing resolution, and overhead reduction.

\section{Performance Evaluation}
In this section, we provide a case study to evaluate the performance of UAV communication and sensing at THz bands. 
We consider a THz-ISAC-UAV system with carrier frequency $0.15$ THz and the transceiver array aperture of the UAV payload is $0.1$ m. 
The number of RF chains is equal to the sum of the user and target number. 
The system bandwidth is set to $3.2$ GHz with the OFDM subcarrier spacing and cyclic prefix duration being $1.5625$ MHz and $0.16$ $\mu \text{s}$, respectively. 
The UAV flies at altitude $20$ m to serve one communication user and sense three targets with RCS being $1$ m$^2$ and the distance of $20$ m, $21$ m, and $23$ m, respectively.
The THz ray-based channel model with a single LoS path is considered with the Rician factor being $30$ dB. 
The tensor method in \cite{Zhang2024TWC} is employed for simultaneous target sensing and channel estimation.
In terms of the benchmark methods for target sensing, we employ the multiple signal classification (MUSIC) method for angle estimation and matching filter (MF) method for range estimation \cite{Liu2020Joint}.
For communication channel estimation, we employ the compressed sensing (CS) method with orthogonal matching pursuit algorithm for performance comparison.
The ISAC performance at the mmWave band is also compared, where the carrier frequency and system bandwidth are set to $30$ GHz and $200$ MHz, respectively.

Fig. 5(a) compares the root mean square error (RMSE) performance of target angle and range estimation for various methods at mmWave and THz bands. 
It can be observed that the tensor method can achieve superior angle and range estimation accuracy compared to conventional MUSIC and MF methods. This improvement is attributed to its ability of better exploiting the multidimensional features of the THz-ISAC channels. In contrast, conventional methods return high estimation errors due to the restricted number of transceiver RF chains in the UAV payload. 
Fig. 5(b) illustrates the achievable rate performance of the THz-ISAC-UAV system. The achievable rate achieved by the tensor method is comparable to that with perfect CCSI under the hybrid beamforming-based transceiver, while the CS method suffers from significant rate loss due to the limited training overhead. 
Besides, we can observe that compared to the mmWave band, the ISAC performance at the THz band shows significant improvement, benefiting from the abundant signal bandwidth and massive arrays of the THz-ISAC-UAV transceiver.  
Fig. 5(c) depicts the RMSE performance in the presence of UAV location uncertainty, where the bias between the actual and preset UAV locations is set to $0.1$ m. The estimation performance of both angle and distance experiences a certain degree of degradation. Note that the target sensing performance degradation at the THz bands is more pronounced compared to that at the mmWave band.

\section{Future Directions}

Despite the advancement of transceiver design and key techniques, there are still several major problems for the research of THz-ISAC-UAV. 
In this section, we will discuss and highlight the potential future research directions.

\subsection{THz-ISAC Performance Metrics and Limits}
The study of THz-ISAC performance metrics and limits can help guide the optimization design of waveforms and systems while meeting communication and sensing requirements.
Existing works on fundamental limits between communication and sensing performance primarily focus on the metrics such as achievable rate, estimation accuracy, transmit beampattern, etc., but generally overlook the energy efficiency-related metrics.  
For the resource-constrained THz-ISAC-UAV transceiver payload with multi-functionality, it is important to incorporate new efficiency-related performance metrics to measure the sensing, or even computing functionality, and provide the basis for the performance analysis of fundamental limits for THz-ISAC. 
Another emerging performance metric in various UAV-enabled applications is the freshness of the received communication or sensing information, particularly for applications demanding information freshness.   
This necessitates the interest in age of information (AoI) performance metrics in communication systems to describe the timeliness of information delivery \cite{Yu2022Age}.
However, from a sensing perspective, the concept of information freshness has not been clearly defined as that for communications. 
How the AoI performance metric can be naturally and efficiently utilized to guarantee the timely information delivery and acquisition requires further investigation for THz-ISAC-UAV.

\subsection{Lightweight Machine Learning-Based Methods}

With the aided of inherent learning capability, machine learning (ML) has been regarded as a powerful tool for addressing challenging problems for 6G wireless communication and sensing applications \cite{Demirhan2023Integrated}.  
For example, one can exploit the received signals as input features to train a convolution neural network, where a nonlinear relationship between the low-dimensional signals and the high-dimensional channels can be established. 
The extracted delay, Doppler, or angle information can be utilized as the extra side knowledge, to proactively predict future beams in a rapidly changing wireless environment. 
Albeit of the tremendous advantages, the data-driven-based methods have not been well tailored for the payload design, which demand low computational delay, storage, and complexity.  
Besides, the high-speed movement of UAV platforms and the ultra-narrow beams of UM-MIMO, present a significant challenge for the available training dataset size and transfer learning capability.  
Therefore, for the UAV-mounted THz-ISAC payloads with limited power budget, how to develop the specific lightweight data-driven or model-driven ML-based methods requires further investigation, particularly under the condition of small-size datasets.

\subsection{Non-Ideal Factors and Interference Mitigation}

Due to the peculiarity of THz bands and UAV platforms, certain non-ideal factors and interferences will be inevitably introduced into UAV transceiver payloads, which may affect the THz-ISAC performance. 
Firstly, THz RF front-end impairments, such as phase noise at the local oscillator and nonlinear distortion of PA, will cause complicated non-linear distortion and fundamentally limit the achievable communication and sensing performance.
Secondly, the synchronization errors between different sensing or communication transceivers, including frequency offset and timing error, will be even exacerbated due to the UAV mobility, especially degrading the THz-ISAC performance with a high requirement on the sensing accuracy.
Thirdly, unpredictable UAV shaking and wave motion caused by wind, as well as the clutter interference, may deteriorate the quality of received signals and target echoes, and cause more frequent switching, significant delays, and adverse impacts on target detection reliability.
Therefore, THz-ISAC systems under rigorous UAV payload constraints necessitate the customized designs for mitigating these imperfections.

\section{Conclusion}
In this article, we have investigated the dominant barriers and key techniques of THz-ISAC-UAV from the transceiver design perspective. 
We first revealed the channel characteristics for UAV-mounted THz communication and sensing.
We then point out the transceiver payload design peculiarities for THz-ISAC-UAV from three aspects of antenna design, RF front end, and baseband signal processing. 
To improve the energy and hardware efficiency, as well as to overcome distance and mobility issues, we shed light on three key technologies, including UM-MIMO-ISAC hybrid beamforming, THz-ISAC waveform design, and C\&S channel state information acquisition, and extensively elaborate their concepts and key issues. 
Finally, potential research directions are highlighted, which may advance the applications of THz-ISAC-UAV for future 6G wireless networks.

% and associated open problems 

% trigger a \newpage just before the given reference
% number - used to balance the columns on the last page
% adjust value as needed - may need to be readjusted if
% the document is modified later
%\IEEEtriggeratref{8}
% The "triggered" command can be changed if desired:
%\IEEEtriggercmd{\enlargethispage{-5in}}

% references section

% can use a bibliography generated by BibTeX as a .bbl file
% BibTeX documentation can be easily obtained at:
% http://mirror.ctan.org/biblio/bibtex/contrib/doc/
% The IEEEtran BibTeX style support page is at:
% http://www.michaelshell.org/tex/ieeetran/bibtex/
%\bibliographystyle{IEEEtran}
% argument is your BibTeX string definitions and bibliography database(s)
%\bibliography{IEEEabrv,../bib/paper}
%
% <OR> manually copy in the resultant .bbl file
% set second argument of \begin to the number of references
% (used to reserve space for the reference number labels box)

\bibliographystyle{IEEEtran}

% that's all folks
\end{document}